\documentclass[useAMS,usenatbib]{mn2e}
\usepackage{graphicx}
\title[Spiral arms in scattered light images of protoplanetary discs]{Spiral arms in scattered light images of protoplanetary discs: Are they the signposts of planets?}

\author[A. Juh\'asz, M. Benisty, A. Pohl, C.P. Dullemond,  C. Dominik and S.-J. Paardekooper]
{
A. Juh\'asz$^{1,2}$\thanks{E-mail:juhasz@ast.cam.ac.uk}, 
M. Benisty$^{3}$, A. Pohl$^{4,5}$, C.P. Dullemond$^{4}$, C. Dominik$^{6}$ and \newauthor S.-J. Paardekooper$^{7,8}$\\
$^{1}$Institute of Astronomy, Madingley Road, Cambridge CB3 OHA, United Kingdom\\
$^{2}$Leiden Observatory, Leiden University, P.O. Box 9513, NL-2300 RA Leiden, The Netherlands\\
$^{3}$University Grenoble Alpes, IPAG, F-38000 Grenoble, France, CNRS, IPAG, F-38000 Grenoble, France\\
$^{4}$Institute for Theoretical Astrophysics, Heidelberg University, Albert-Ueberle-Strasse 2, D-69120 Heidelberg, Germany\\
$^{5}$Max-Planck-Insitute for Astronomy, K\"onigstuhl 17, D-69117 Heidelberg, Germany\\
$^{6}$Anton Pannekoek Institute for Astronomy, University of Amsterdam, Postbus 94249, NL1090 GE Amsterdam, The Netherlands\\
$^{7}$Astronomy Unit, School of Physics and Astronomy, Queen Mary, University of London, Mile End Road, London E1 4NS, UK\\
$^{8}$DAMTP, University of Cambridge, Wilberforce Road, Cambridge CB3 0WA, UK
}

\begin{document}

%\date{Accepted 1988 December 15. Received 1988 December 14; in original form 1988 October 11}

%\pagerange{\pageref{firstpage}--\pageref{lastpage}} \pubyear{2002}

\maketitle

\label{firstpage}

\begin{abstract}
One of the striking discoveries of protoplanetary disc research in recent years are the spiral arms seen in 
several transitional discs in polarised scattered light. An interesting interpretation of the observed spiral features
is that they are density waves launched by one or more embedded (proto-)planets in the disc. In this paper
we investigate whether planets can be held responsible for the excitation mechanism of the observed spirals.
We use locally isothermal hydrodynamic simulations as well as analytic formulae to model the spiral waves 
launched by planets. Then {\it H}-band scattered light images are calculated using a 3D continuum radiative
transfer code to study the effect of surface density and pressure scale height perturbation on the detectability of the spirals. 
We find that a relative change of $\sim$3.5 in the surface density ($\delta\Sigma/\Sigma$) is required for the spirals to be detected with
current telescopes in the near-infrared for sources at the distance of typical star-forming regions (140\,pc).
This value is a factor of eight higher than what is seen in hydrodynamic simulations. 
We also find that a relative change of only 0.2 in pressure scale height is sufficient to create detectable signatures
under the same conditions. Therefore, we suggest that the spiral arms observed to date in 
protoplanetary discs are the results of changes in the vertical structure of the disc (e.g. pressure scale height perturbation) 
instead of surface density perturbations.

\end{abstract}


\section{Introduction}
\label{sec:introduction}
Today, more than a thousand extrasolar planets have been detected around main-sequence stars and these planetary systems show a great diversity in their architectures \citep{batalha_2013}. However, none of the claimed detections of planetary candidates in their native circumstellar disc (e.g., \citealt{kraus_2012, huelamo_2011, quanz_2013}) have been confirmed so far. The direct observation of a (proto-)planet in a disc is difficult due to the large brightness contrast with the disc/star system that outshines it. Nonetheless, a massive planet could be revealed by indirect signatures of its interaction with the disc. Planets more massive than Jupiter gravitationally interact with the surrounding gas, and carve a gap or a cavity in the disc \citep{crida_2006}. As a result of planet-disc interactions, a number of non-axisymmetric features can appear, such as spiral arms, warps \citep{facchini_2014} or an overall spatial discrepancy between small and large dust grains \citep{pinilla_2012, de_juan_ovelar_2013}. 

Hydrodynamic processes that do not involve planets can also lead to vortices or spirals (see \citealt{turner_2014} for a review). 
Asymmetric features can result from gravitational instability \citep{lodato_2004, boley_2006, rice_2006a, baruteau_2011} or tidal interaction with an external companion \citep{papaloizou_2001}. Recent numerical simulations also show that non ideal MHD can spontaneously lead to such non-axisymmetric structures 
\citep{kunz_2013}.

Asymmetric features have been recently observed in transition discs, which are thought to be at an advanced evolutionary 
stage. They show a dip in their infrared (IR) spectral energy distribution (SED), indicative of  a dust-depleted cavity (see \citealt{espaillat_2014} for a review), that can sometimes be imaged  \citep{andrews_2011}. High angular resolution images of their outer 
regions have indeed shown a variety of asymmetries in the sub-millimetre continuum tracing cold material (e.g., 
\citealt{casassus_2013, van_der_marel_2013}), and in the near infrared (NIR) through scattered light that traces small hot dust in the disc surface layers \citep{fukagawa_2006, grady_2013}. For example, in the case of SAO206462, two spiral arms can be traced up from 28 to 140~AU in NIR scattered light with a polarised intensity about 30 per cent larger than the background disc \citep{muto_2012, garufi_2013}. In HD142527, six spiral arms can be traced in scattered light which start from the edge of its 142~AU-radius gap  \citep{avenhaus_2014b}. Sub-millimetre ALMA observations indicate the presence of three CO arms extending up to 670 AU, but only one is a radio counterpart of the IR arms \citep{christiaens_2014}. The only previous detection of spiral features in such wavelength range was in AB Aur \citep{pietu_2005} but the spirals were found to be in counter rotation with the disc, indicating a likely origin in the late envelope infall.  What is causing the gap or cavity in transition discs is still debated, but at least in some cases, it can be due to the dynamical interaction of one or multiple planets with the disc \citep{zhu_2011, dodson_robinson_2011}. On the one hand, photoevaporation models fail to reproduce the properties of  transition discs  with high accretion \citep{owen_2012} or NIR excess (such as HD100546, \citealt{tatulli_2011}), while on the other hand, grain growth models could reproduce the IR dip seen in the SED, but not the millimetre images \citep{birnstiel_2012}. Thus, the presence of a planet has been  often invoked to explain the observed asymmetries \citep{muto_2012, garufi_2013, van_der_marel_2013}.

While the presence of a planetary mass companion is a very intriguing explanation for the observed asymmetries, especially for spiral arms, no quantitative comparison
between observation and theory has been done so far. Observational studies compared the morphology of spiral arms to those seen in 
hydrodynamic simulations, e.g. by fitting the spiral wake \citep{muto_2012}.  However, it has not yet been investigated whether the 
observed amplitude of the spirals agrees with that in theoretical calculations. In this paper we study the detectability of planet-induced spiral density waves 
in scattered light observations.  Using a combination of 2D hydrodynamic and 3D radiative transfer codes we test whether the spirals observed so far can
be explained by planet-induced spiral density waves and we also make predictions for future observations with next generation instruments.

\section{Model setups}
\label{sec:hydro_simulations}
\subsection{Hydrodynamic simulations}
\label{subsec:hydro_model_setup}

We use the 2D hydrodynamic code \textsc{fargo} \citep{masset_2000} to study the shape and amplitude of the spiral density
waves launched by an embedded planet. The input parameters of the simulations are summarised in Table\,\ref{tab:fargo_params}.
For all simulations with planet-to-star mass ratio of $M_{\rm pl}\leq10^{-3}M_{\star}$ we use the same spatial grid. 
However, for the most massive case ($M_{\rm pl}=10^{-2}M_{\star}$) we extend the outer boundary of the computational
grid in order to minimise the effect of the outer boundary conditions on the structure of the disc around the gap \citep{kley_2006}. 
The surface density distribution is assumed to be $\Sigma = \Sigma_0  (R/R_{\rm in})^{-0.5}$ where $\Sigma_0=0.003$. 
Assuming  $R_{\rm pl}=25$\,AU, the gas mass in our setup  between 5.0\,AU and 100\,AU is 0.01\,$M_\star$.  

A single planet is placed in the disc on a fixed circular orbit with a radius of unity in dimensionless units. In the reference model the 
mass of the planet is $10^{-3}M_\star$. The accretion onto the planet is switched off and the simulations are run for 1000 planetary 
orbits. We vary the mass of the planet and the disc aspect ratio to study their effect on the disc structure and on the density perturbation 
induced by the planet. 

The simulations are locally isothermal not taking into account heating and cooling of the disc, which is a reasonable assumption
if the disc is optically thin to its own radiation resulting in efficient cooling.  Apart from surface density perturbations, 
temperature perturbations are also expected along the spirals due to the presence of shocks along the spiral. 
We study the effect of pressure scale height perturbations with analytic models in Sec\,\ref{subsubsec:hpperturb_analytic}, but the full treatment of heating 
and cooling in the hydrodynamic simulations is the topic of a forthcoming paper. 

\begin{table}
	\caption[\textsc{fargo} input parameters]{\textsc{fargo} input parameters}
	\label{tab:fargo_params}
	\begin{tabular}[h]{@{}lcc}
		\hline\hline \textbf{Parameter} & \textbf{Value} & \textbf{Value} \\
		\hline
		$R_{\mathrm{in}}$ & 0.2\,$R_{\mathrm{pl}}$& 0.2\,$R_{\mathrm{pl}}$\\
		$R_{\mathrm{out}}$ & 4.0\,$R_{\mathrm{pl}}$& 7.0\,$R_{\mathrm{pl}}$\\
		$N_{\mathrm{r}}$ & 384 & 768\\
		$N_{\mathbf{\varphi}}$ & 512 & 1024\\
		Inner Boundary & non-reflecting & non-reflecting\\
		Outer Boundary & closed & closed\\
		$\mathbf{\alpha_{\rm visc}}$ & 10$^{-3}$& 10$^{-3}$\\
		$H_{p}/R$ & 0.05, 0.1 & 0.05, 0.1\\
		$M_{\mathrm{pl}}/M_{\star}$ & $\lbrace  6.25\cdot10^{-5}, 10^{-3} \rbrace$ & $10^{-2}$ \\
		\hline  
	\end{tabular}
	
	\medskip
	\textbf{Notes.} \small{All parameters are given in \textsc{fargo} units, i.e. with respect to the planet's radius $R_{\mathrm{pl}} = 25$\,AU and primary mass $M_{\star} = 1.0M_{\odot}$. In the reference model the mass of the planet is $10^{-3}M_{\star}$.}
\end{table}  

\begin{figure*}
	\includegraphics[width=16.0cm]{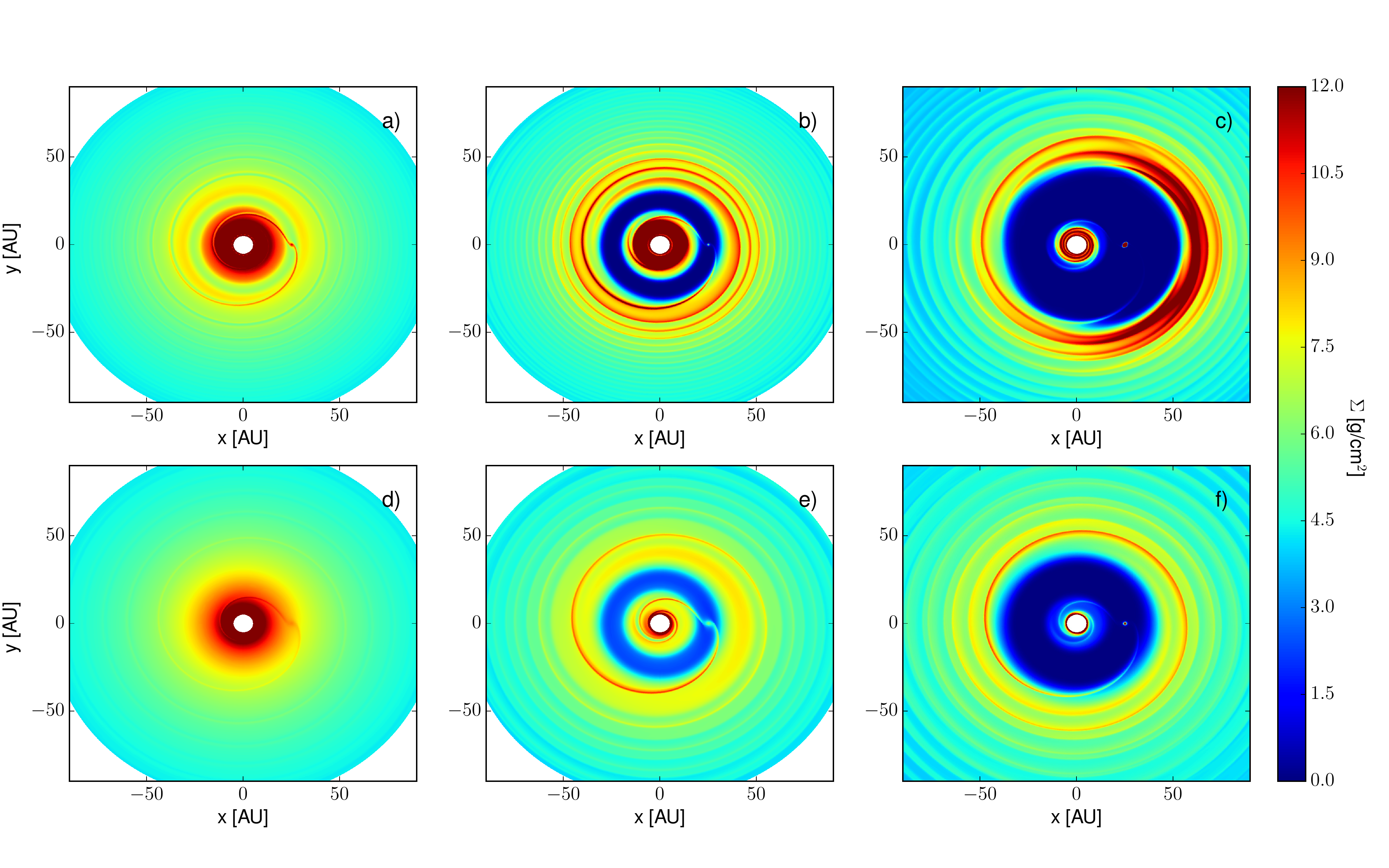}
	\caption{Surface density structure of the disc with an embedded planet for a planet-to-star mass ratio of $6.25\cdot10^{-5}$ ({\it a,d}), $10^{-3}$ ({\it b,e}) and
 of $10^{-2}$ ({\it c,f}). The upper row shows the surface density for an aspect ratio of $H_{p}/R=0.05$ while the bottom row for $H_{p}/R=0.1$. The 
 strength and the number of the spirals increases for higher planetary masses and decreases for higher aspect ratio. The spirals become also more open
 for higher aspect ratios.} 
	\label{fig:FARGO_comp_low_highmass}
\end{figure*}

\subsection{Radiative transfer}
\label{subsec:RT_model_assumptions}
We use the 3D radiative transfer code \textsc{radmc-3d}\footnote{http://www.ita.uni-heidelberg.de/~dullemond/software/radmc-3d/}  to calculate near infrared scattered light images. In the radiative transfer
calculations we use a spherical mesh ($r,\theta,\phi$), where the radial and azimuthal grid are taken to be the same as in the 
hydrodynamical simulations. In the poloidal direction we use $N_\theta=250$ grid cells in total. 
To ensure that we resolve the upper layers of the disc, which is responsible for the observed scattered light, we
place $N_\theta=\{10,100,30,100,10\}$ points to the $\left[0,\pi/2-\theta_1\right]$,  
$\left[\pi/2-\theta_1,\pi/2-\theta_2\right]$, $\left[\pi/2-\theta_2,\pi/2+\theta_2\right]$, $\left[\pi/2+\theta_2,\pi/2+\theta_1\right]$, 
$\left[\pi/2+\theta_1],\pi\right]$ intervals, respectively, where $\theta_1 = 9H_{p}/R$ and  $\theta_2 = 3H_{p}/R$, with
$H_{p}$ being the pressure scale height in the disc.

The density structure of the disc in the radiative transfer calculations is given as
\begin{equation}
\rho(R,z,\phi) = \frac{\Sigma(R,\phi)}{\sqrt{2\pi}H_{p}(R )}\exp{\left(-\frac{z^2}{2H_{p}(R )^2}\right)}
\label{eq:disc_density}
\end{equation}
where $\Sigma(r,\phi)$ is the surface mass density,  $H_p(R )$ is the pressure scale height, $R=r\sin{\theta}$ and $z=r\cos{\theta}$.
The radial variation of the aspect ratio of the disc is described by a power-law,
\begin{equation}
\frac{H_{p}(R )}{R} = \frac{H_{p}(R_{\rm pl})}{R_{\rm pl}} R^\zeta,
\label{eq:flaring_index}
\end{equation}
where $R_{\rm pl}$ is the radius of the planetary orbit and $\zeta$ is the flaring index. 
The aspect ratio and the flaring index was taken to be the same as in the hydrodynamic simulations. 

For the radiation source we assume a central star with parameters representative for that of a Herbig Ae star, $T_{\rm eff}=9500$\,K, 
$R_\star=2.5\,R_\odot$, $M_\star=2.0\,M_\odot$. For simplicity the stellar radiation field is described by blackbody emission.

The dust in the disc consists of 0.1\,$\mu$m sized silicate grains, whose scattering matrix is calculated from the optical 
constants of astronomical silicates \citep{weingartner_2001} using Mie-theory. In reality dust grains are likely
to have a wide size distribution, which we neglected here. Dust grains which are small compared to the wavelength scatter
photons uniformly, while large grains have strongly forward peaking scattering phase function. This means that for discs viewed at
low inclination angles large grains  scatter stellar photons mostly into the disc, and only a small fraction toward the observer, making the disc faint in 
scattered light (see \citealt{mulders_2013}). Therefore, as long as  dust properties are uniform in the disc or at least
there is no steep gradient in the dust size distribution, the main effect of a wider grain size distribution is to scale the total 
scattered flux of the disc. To maximise the computational efficiency, i.e. to have the largest number of photons scattered towards the observer
also for low disc inclinations, we choose to use a single grain size of 0.1\,$\mu$m.

We also assumed that the he dust-to-gas ratio is uniform in the disc, thus we can model the dust distribution by simply scaling the gas 
distribution we calculate with FARGO. While the dust-to-gas ratio is known to get enhanced in the spirals excited by gravitational instability
(see e.g., \citealt{rice_2004}), planetet-induced spirals are unlikely to trap dust particles. Spiral waves launched by a planet are co-rotating with 
the planet, i.e. they will move with respect to the background gas flow. Dust particles can therefore only be trapped in the spirals if the
dust accumulation time-scale is shorter than the time-scale for the spirals to pass by. As it was shown by \citet{birnstiel_2013} the dust accumulation 
time-scale is hundred times longer than the local orbital period even in the most optimistic case, while the time-scale of the spiral pass 
by is only few orbital time-scale of the planet even very close to the planet. Thus particle trapping is only possible for structures which are either
close to co-rotation with the local gas-flow, like e.g. spiral arms in gravitationally unstable discs, or for axisymmetric structures, like the pressure 
bump at the outer edge of the gap. Moreover, dust trapping works for only large grains which are at least partially decoupled from the gas 
(see e.g. \citealt{lyra_2009, pinilla_2012,birnstiel_2013}), while the micron- and sub-micron sized dust particles, responsible for near-infrared 
scattered light, are rather well coupled to the gas. Thus we think that the assumption of uniform dust-to-gas ratio in the disc does not affect the results of this paper.

The radiative transfer calculations contain two steps. First, the temperature structure of the dust is determined in a thermal Monte Carlo 
simulations then images are calculated including the full treatment of polarisation. For the thermal Monte Carlo simulations we use 
$1.6\cdot 10^8$ photon packages while the {\it H}-band images are calculated using $6\cdot 10^7$ photon packages. The square images are 
calculated at 1.65\,$\mu$m, at an inclination angle of 10$^\circ$ with 700 pixels along the edge, covering a linear scale of 250AU. This 
corresponds to a spatial resolution of 0.357AU/pixel. To convert linear scales to angular coordinates we assume the source to be at 140\,pc. 
\textsc{radmc-3d} calculates images in the full Stokes vector (I,Q,U,V) from which the polarised intensity can be calculated as 
${\rm PI} = \sqrt{{\rm Q}^2+{\rm U}^2}$. 

These images represent the true sky brightness distribution, i.e. virtually at infinite spatial resolution. In reality the observed
images are the convolution of the true sky brightness distribution with the point-spread-function (PSF) of the telescope. 
To simulate the observations the images need to be convolved with the telescope PSF. We assume the simplest 
PSF model, a Gaussian, which is a good representation of the Airy-disc, the core of a diffraction limited PSF. First, we
investigate whether or not the already observed spirals in transitional discs can be explained with planet-induced
density waves. For this purpose we convolve the images with a PSF whose full-width at half maximum (FWHM) is 
taken to be 0.06", representative for the observations  with HiCIAO/SUBARU presented in \citet{muto_2012}. 
Then we make predictions for the observability of spirals in future observations with next generation instruments (e.g.
SPHERE/VLT) and telescopes (E-ELT) by changing the size of the PSF the images are convolved with.

\section{Results from hydrodynamic simulations}

\subsection{Morphology of the spirals}
\label{subsec:hydro_spiral_morphology}

Low-mass planets which do not open gaps are known to drive a single one-armed spiral forming
as a constructive resonance between waves launched at Lindblad-resonances both inside and outside of the planetary orbit 
 \citep{ogilvie_2002}. However, for massive planets which open gaps, the wave propagation is highly non-linear the 
 morphology of the spirals might be different. We studied the effect of the planet mass and disc aspect ratio on the morphology of the spirals. 

To study the effect of the planet mass on the strength of the spiral perturbation we run simulations with various planet masses 
($M_{\rm pl}/M_\star=\{6.25\cdot 10^{-5}, 10^{-3}, 10^{-2}$\}). 
In Fig.\,\ref{fig:FARGO_comp_low_highmass} we show the surface density distribution of the disc for three different planet masses 
after 1000 planetary orbits. The spiral density waves are clearly visible in all cases. However, the structure of the spirals in the outer disc
changes with the planetary mass. For a low mass planet ($M_{\rm pl}/M_\star=6.25\cdot 10^{-5}$) there is a well defined, single spiral 
arm in the outer disc (see Fig.\,\ref{fig:FARGO_comp_low_highmass}\,a). For a more massive planet ($M_{\rm pl}/M_\star=10^{-3}$), that 
opens a gap, there are two spiral arms. The two arms of the spiral are, however, not completely symmetric. 
The one that goes through the planet position has a smaller radial width and higher amplitude than 
the secondary spiral that is shifted in azimuth compared to the planet position by about $180^{\circ}$  (see Fig.\,\ref{fig:FARGO_comp_low_highmass}\,b). 
For the most massive planet in our simulation ($M_{\rm pl}/M_\star=10^{-2}$) the gap becomes eccentric and several spiral arms become visible
(see Fig.\,\ref{fig:FARGO_comp_low_highmass}\,c). The strength and sharpness of the secondary spiral arm change smoothly 
with the planet mass.

\begin{figure*}
\includegraphics[width=16.5cm]{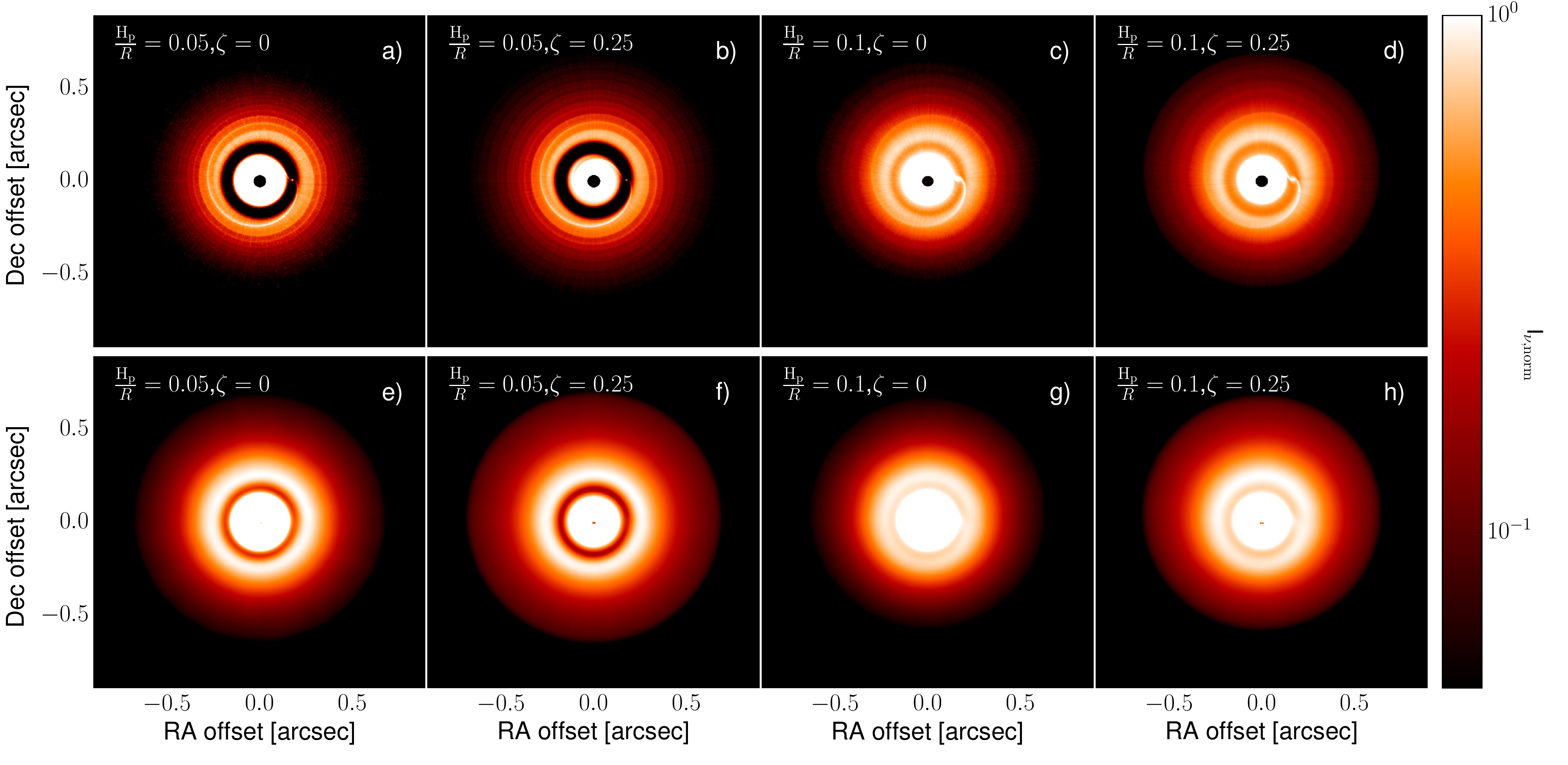}
\caption{Simulated {\it H}-band polarised intensity images with the surface density given by hydrodynamic simulations with a planet to star mass ratio of 0.001 for different 
values of the disc aspect ratio ($H_p/R$) and flare index ($\zeta$). 
The images are normalised to the highest intensity outside of a 0.18" radius (25.2\,AU at 140\,pc distance) from the 
central star. The top row shows images at original resolution as calculated by \textsc{radmc-3d}, while the bottom row shows the images convolved with a circular 
Gaussian beam with an FWHM of 0.06".  While spiral arms are clearly visible in the original, quasi-infinite resolution ({\it Top row}), no spiral arm is visible after convolving the images with a 0.06" PSF ({\it Bottom row}) in any of the images.} 
\label{fig:image_FARGO_allsim_1mjup}
\end{figure*}

Variations in the aspect ratio of the disc, i.e. the ratio of the pressure scale height and the radius, changes the temperature and the sound speed 
in the disc, which in turn affects the propagation of sound waves through the disc. 
The angle between the radial direction and the tangent of the spiral, the complement of the pitch angle, is inversely proportional to the sound speed 
and the pressure scale height \citep{rafikov_2002a}. Thus increasing aspect ratio increases the pitch angle, making the spiral pattern more open  
(see Fig.\,\ref{fig:FARGO_comp_low_highmass} top row and bottom row). The strength of the spirals, which are offset in azimuth
with respect to the position of the planet, decreases with increasing aspect ratio. As can be seen in Fig.\,\ref{fig:FARGO_comp_low_highmass}b the 
secondary spiral is clearly visible for an aspect ratio of $H_{p}/R=0.05$ in case of a planet mass of $10^{-3}M_{\star}$. If the aspect ratio is 
increased to $H_{p}/R=0.1$ the secondary spiral weakens to a barely recognisable level. By increasing the aspect ratio from 0.05 to 0.1 for a planet
mass of $10^{-2}M_{\star}$ the multi-armed  spiral structure reduces to a two-armed spiral, similar to the case of a $10^{-3}M_{\star}$ planet with 
$H_{p}/R=0.05$. 

The dependence of the pitch angle of the spirals on the local pressure scale height of the disc provides an important link
	between the azimuthal and the vertical structure of the disc. By measuring the pitch angle of the spirals, or fitting the spiral wake, 
	one can estimate the pressure scale height of the disc and can get some clues about the vertical structure of the disc. The pressure
	scale height and its radial variation (i.e. the flaring index) are on the other hand setting also the total luminosity of the disc, 
	the slope of the spectral energy distribution (SED)  from near- to far-infrared wavelengths, as well as the absolute surface brightness distribution 
	of the background disc. The combination of these two constraints (pitch angle of the spirals, luminosity of the disc) can be a powerful
	tool to test the origin of the spirals.  In case the spiral is driven by an embedded planet, the simultaneous fitting of the SED and the spiral wake
	should  result in the same physically plausible curve for the pressure scale height as a function of radius $H_{\rm p} (R)$.

\subsection{Number of spiral arms}
\label{subsec:hydro_spiral_number}
The perturbation induced by low-mass planets, for which $M_\mathrm{pl}/M_* \ll (H_{p}/R)^3$, can be calculated by decomposing the potential of the planet into 
Fourier harmonics azimuthally and solving the resulting ordinary differential equations for the linear perturbations numerically (e.g. Korycansky \& Pollack 1993). The 
resulting surface density perturbation is a one-armed spiral density wave. It was shown by \citet{ogilvie_2002} that this wave results from interference between the 
different azimuthal modes, and that as long as a range of azimuthal wave numbers $m$ is present, and there are no special selection rules on $m$, there is only one 
value of $\varphi$ at any radial location where constructive interference can occur; hence a one-armed spiral. 

However, when  $M_\mathrm{pl}/M_*$ approaches $(H_{p}/R)^3$, linear theory breaks down \citep{korycansky_1996}. One can probe this regime by 
going beyond linear theory, and include second-order perturbations \citep{artymowicz_1992}. These perturbations do not couple to the planet potential directly, since the potential is completely 
specified at linear order, but instead to quadratic terms in for example the linear perturbation in radial velocity. Since these quadratic terms have twice the azimuthal wave 
number of the corresponding linear terms, the second-order solution will only involve even numbers of $m$. This selection rule allows for interference at \emph{two} 
values of $\varphi$ at any radial location: one at the original location of the one-armed spiral, and one shifted by $\pi$. 
A second effect that sets in as $M_\mathrm{pl}/M_\star$ approaches $(H_{p}/R)^3$ is that the width of the horseshoe region becomes 
comparable to the scale height of the disc \citep{paardekooper_2009}. 
This means that the 
highest-order resonances, located approximately one scale height away from the planet, can no longer 
contribute to the wake. In the limit of $M_\mathrm{pl}/M_\star \gg (H_{p}/R)^3$ the $m=2$ becomes 
dominant, resulting again in a two-armed spiral. We leave a detailed description of this mechanism 
to a forthcoming paper (Paardekooper et al., in prep). The shifted spiral arm can be clearly identified for example in Fig.\,\ref{fig:FARGO_comp_low_highmass}\,b. 

This second spiral can only be comparable in magnitude to the original spiral wave if $M_\mathrm{pl}/M_* \sim (H_{p}/R)^3$. For smaller planets, only a one-
armed spiral will be visible. Note that this criterion is similar to the thermal criterion for gap opening (e.g. \citealt{crida_2006}). 
We therefore expect only gap-opening planets 
to exhibit two spiral arms, which is confirmed by Fig\,\ref{fig:FARGO_comp_low_highmass}. For $M_\mathrm{pl}/M_* \gg (H_{p}/R)^3$, additional effects kick in. For example, it is well-known that for 
high enough planet masses the outer disc can become eccentric \citep{papaloizou_2001, kley_2006}. The eccentricity of the disc comes with its own surface 
density perturbations that interfere with the two spirals (see Fig.\,\ref{fig:FARGO_comp_low_highmass}\,c)).

 \begin{figure*}
	\includegraphics[width=5.8cm]{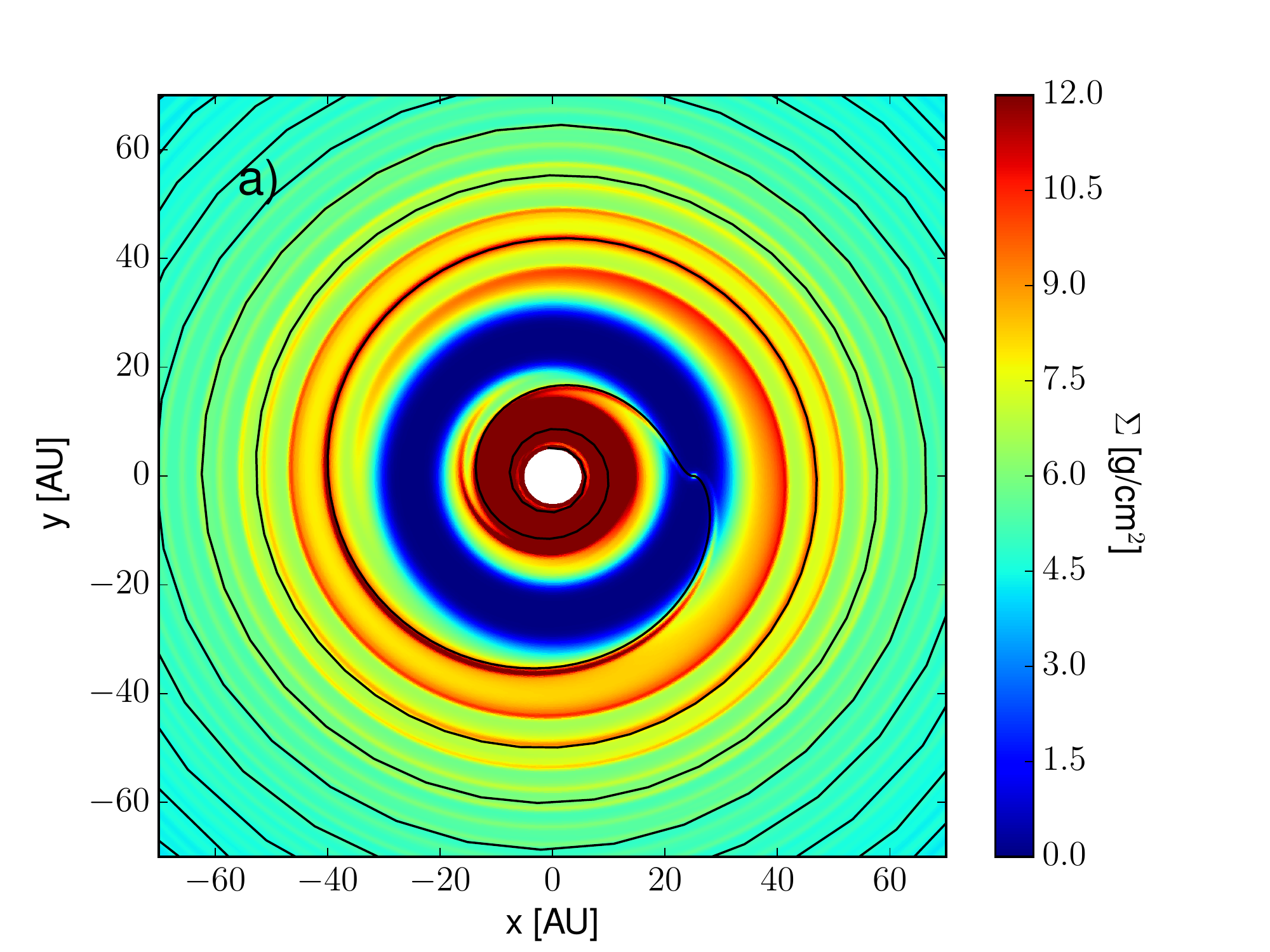}
	\includegraphics[width=5.8cm]{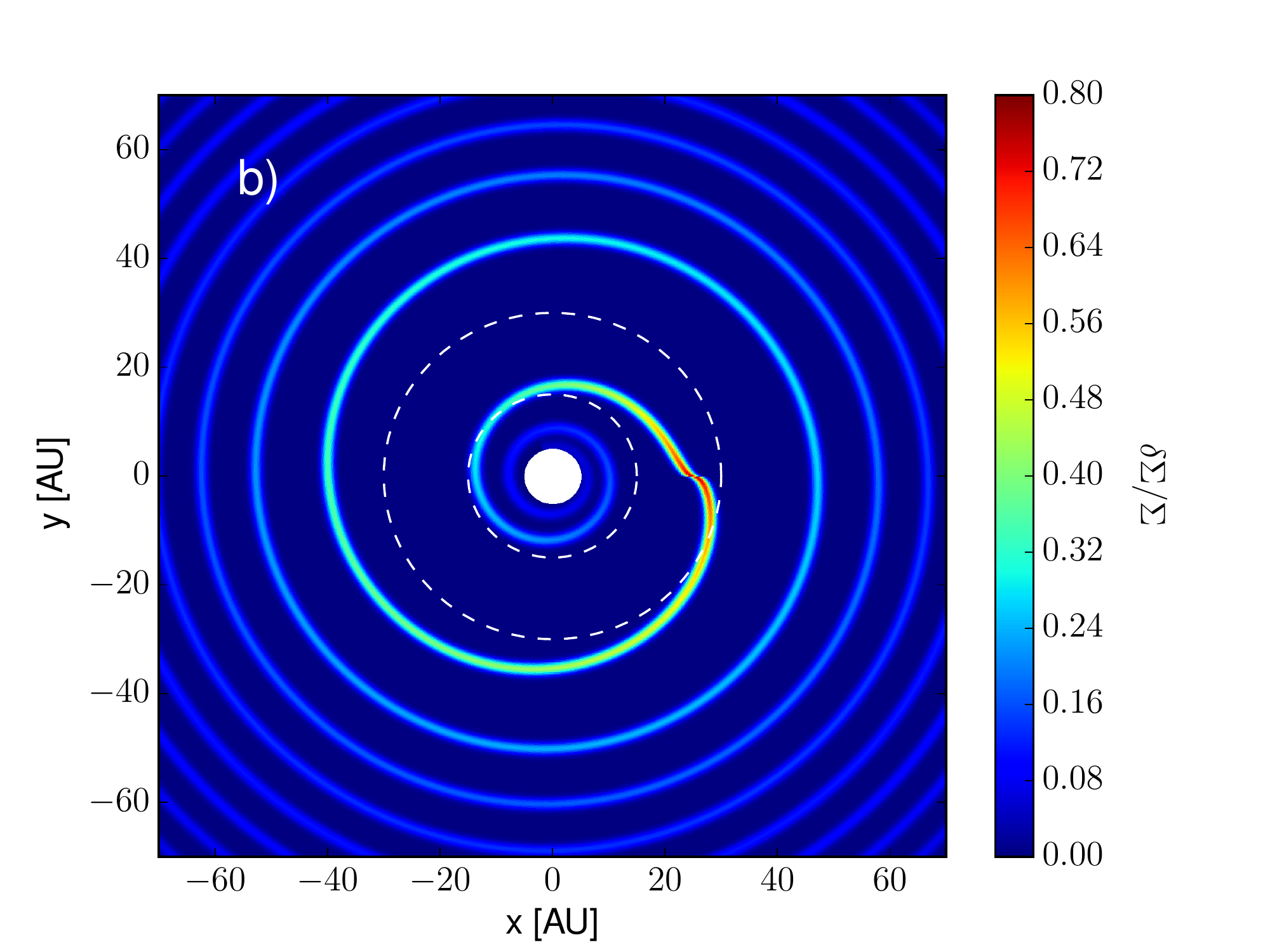}
	\includegraphics[width=5.8cm]{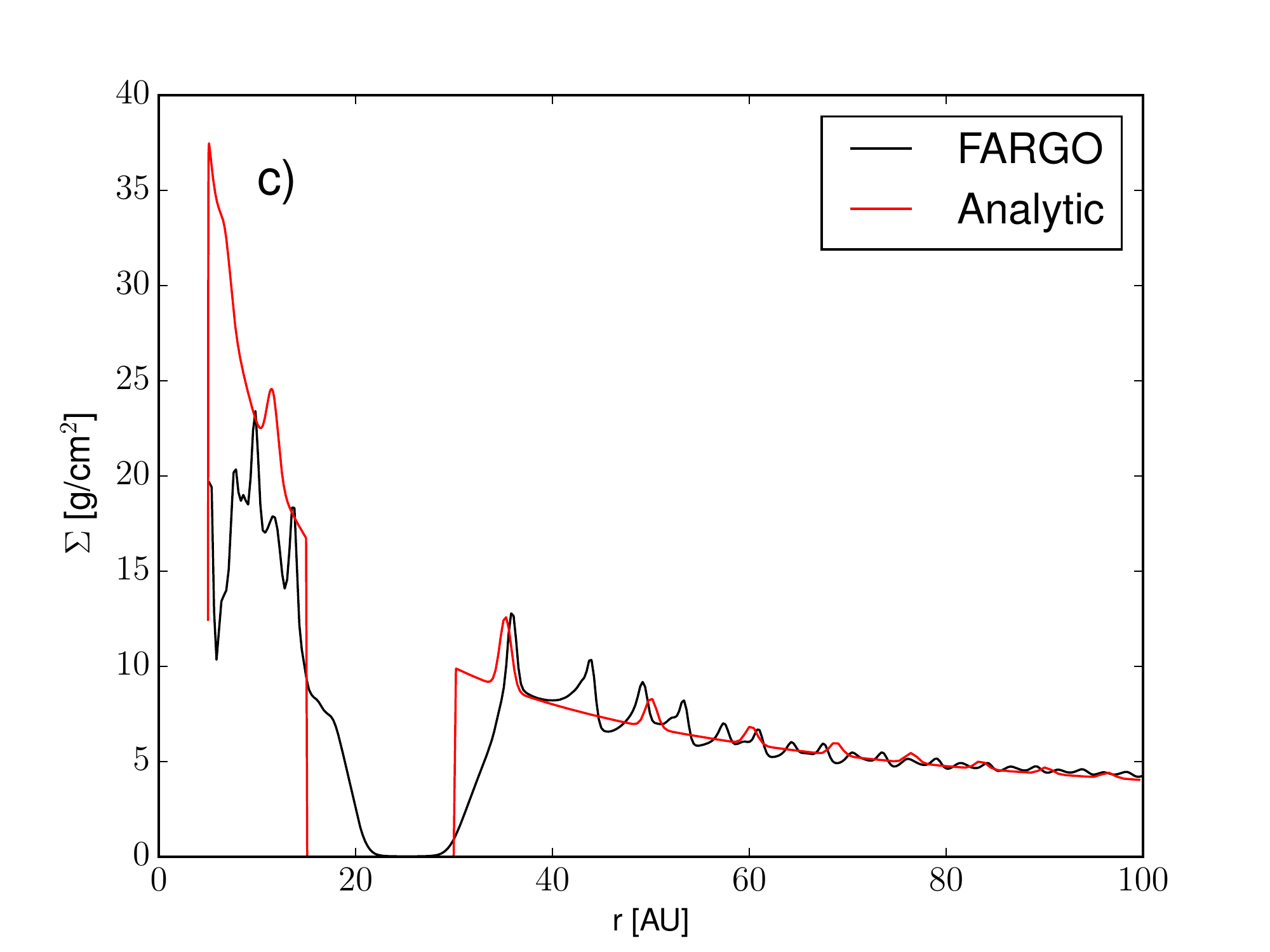}
	\caption{{\it a)} Surface density distribution in the reference hydrodynamical model after 1000 planetary orbits. The black line shows the 
	analytic fit to the spiral wake. {\it b)}  Relative surface density perturbation in the analytic model with a spiral perturbation fitted to the reference 
	hydrodynamical model. The white dashed lines mark the inner and outer radius of the applied gap. {\it c)} Surface density as a function of radius along the 
	$\phi=270^\circ$ azimuth angle (i.e. along the negative y-axis).}
\label{fig:FARGO_fit_perturbation}
\end{figure*}

\subsection{Images from hydrodynamic simulations}
\label{subsec:sdperturb}

We use the surface density distribution ($\Sigma (R,\phi)$) calculated by \textsc{fargo} to describe the disc structure
in Eq.\,\ref{eq:disc_density}-\ref{eq:flaring_index}.  The calculated scattered light images in {\it H}-band polarised intensity for 
a planet-to-star mass ratio or 0.001 are presented in Fig.\,\ref{fig:image_FARGO_allsim_1mjup}{\it a,b,c,d}. 
Since we are interested in the observability of the spirals in the outer disc we normalise the images to the
highest surface brightness outside of the planetary orbit (25\,AU or $\sim$0.179" at 140\,pc distance) and we set the dynamic
range of the images to 20 similar to the images of HD\,135344B presented by \citet{muto_2012}.

The spiral arms are clearly visible in the images calculated by \textsc{radmc-3d} (see Fig.\,\ref{fig:image_FARGO_allsim_1mjup}{\it a,b,c,d}).
The contrast between the spiral and the background disc is higher for lower aspect ratio (see Fig.\,\ref{fig:image_FARGO_allsim_1mjup}{\it a,b}), 
than for a larger one (see Fig.\,\ref{fig:image_FARGO_allsim_1mjup}{\it c,d}), as expected on the basis of the surface density perturbation.
Changing the flaring index affects the brightness of the disc
in the outer regions, but does not affect the visibility of the spirals. The increase of the pressure scale height increases the pitch-angle, i.e. the
tightness,  of the spirals making them more open. However, the increase of the pressure scale height also decreases the amplitude
of the spirals making them even harder to observe. 

As can be seen in  Fig.\,\ref{fig:image_FARGO_allsim_1mjup}{\it e,f,g,h} after convolving the images with a 0.06" PSF, representative for 
the reported observations of transitional discs with 8--10\,m-class 
telescope in the {\it H}-band, the spiral features are not visible anymore. This is true even for the case of a
very massive planetary companion with a planet-to-star mass ratio of 0.01. Therefore it seems to be very unlikely that
the spiral arms observed to date in transitional discs in scattered light are caused by pure surface density perturbations.

\section{Results from analytic models}

\begin{table}
	\caption[Parameters of the analytic models]{
	Parameters of the analytic spiral perturbation as it is fitted to surface density perturbation in 
	the reference model with a planet mass of $M_{\rm pl} = 10^{-3}M_\star$ and aspect ratio of $H_{p}/R=0.05$.}
	\label{tab:analytic_spiral_fitting}
	\begin{tabular}[h]{@{}lc@{}}
		\hline\hline \textbf{Parameter} & \textbf{Value}  \\
		\hline
		 $\alpha$ &  $1.5$  \\
		$\beta$ &  0.51 \\ 
		$R_{\rm pl}$ & 25\,AU   \\
		$\phi_{\rm pl}$  & $0^\circ$  \\
		$h_{\rm pl}$ & 0.067  \\
		$\Sigma_0$ &  4.04 cm$^2$/g\\
		$p$ &  -0.7   \\
		$A$ &  0.8 \\
		$q$   & -1.7  \\
		$\sigma_{\rm sp}$ &  0.8\,AU\\
		\hline  
	\end{tabular}

\end{table}

 \subsection{Analytic model for spiral perturbation}
\label{sec:analytic_models}
Since the spirals are not visible in the PSF-convolved images we calculated based on the hydrodynamic simulations, we
use the analytic models to test the contrast/amplitude requirement for the observability of the spirals.
Our model is based on the WKB solution for propagation of planet-induced density waves in power-law discs  given by \citet{rafikov_2002a}. We assume that any perturbed variable in the disc can be expressed in the form

\begin{equation}
X(R,\phi) = X(R)\cdot\left(1 + \frac{\delta X(R,\phi)}{X(R)}\right).
\label{eq:pert_eq1}
\end{equation}

The first term on the right hand side describes the symmetric unperturbed part, that is assumed to have the form of a power-law
$X(r)=X_0(R/R_{\rm pl})^{p}$ with $R_{\rm pl}$ being the orbital radius of the planet. $\delta X(R,\phi)$ contains the non-axisymmetric
perturbation caused by the planet. The spiral perturbation function $\delta X(R,\phi)$ is described 
by a Gaussian in the radial direction around the centre of the spiral wake  whose amplitude decreases with the radial
distance from the star
\begin{equation}
\delta X(R,\phi) =A\cdot \left(\frac{R}{R_{\rm pl}}\right)^{{\rm sgn}(R-R_{\rm pl})q}  \exp{\left( -\frac{(R-R_0(\phi))^2}{\sigma_{\rm sp}^2} \right)}
\label{eq:pert_eq2}
\end{equation}
The constant $A$ sets the amplitude of the perturbation at the position of the planet, $\sigma_{\rm sp}$ is the 
radial width of the spiral and $R_0(\phi)$ describes the spiral wake. We calculate $R_0(\phi)$ from the wake equation of 
\citet{rafikov_2002a} used also by \citet{muto_2012}
 
 {\small
 \begin{eqnarray}
\nonumber \phi(R) &&= \phi_{\rm pl} - \\
 &&\nonumber \frac{{\rm sgn}(R-R_{\rm pl})}{h_{\rm pl}} \left(\frac{R}{R_{\rm pl}} \right)^{1+\beta} \left\{\frac{1}{1+\beta} - \frac{1}{1-\alpha+\beta}\left(\frac{R}{R_{\rm pl}}\right)^{-\alpha}\right\} \\
 &&+ \frac{{\rm sgn}(R-R_{\rm pl})}{h_{\rm pl}} \left( \frac{1}{1+\beta} - \frac{1}{1-\alpha+\beta}\right)
\label{eq:sp_wake_muto}
\end{eqnarray}}

where $R_{\rm pl}$ is the radius of the planetary orbit, $\alpha$ is the power exponent of the rotation angular frequency ($\Omega(r )\propto r^{-\alpha}$), $\beta$
is the power exponent of the radial distribution of the sound speed ($c_{\rm s} \propto r^{-\beta}$), $h_{\rm pl} = H_{p}(R_{\rm pl})/R_{\rm pl}$ is the aspect ratio of the disc at $R_{\rm pl}$ and $\phi_{\rm pl}$ is the azimuthal coordinate of the planet. 

Our goal is to derive a lower limit for the relative perturbation amplitude ($\delta X(r,\phi)/X(r)$)  required to detect the spirals in scattered light observations and
compare that to the numerical simulations. To keep our analytic model as realistic as possible we fit the 2D surface density distribution of the disc in our reference
hydrodynamic simulation with Eq.\,\ref{eq:pert_eq1}--\ref{eq:sp_wake_muto}. This way we can "calibrate" the shape of the perturbation function.

\begin{figure*}
\includegraphics[width=16.5cm]{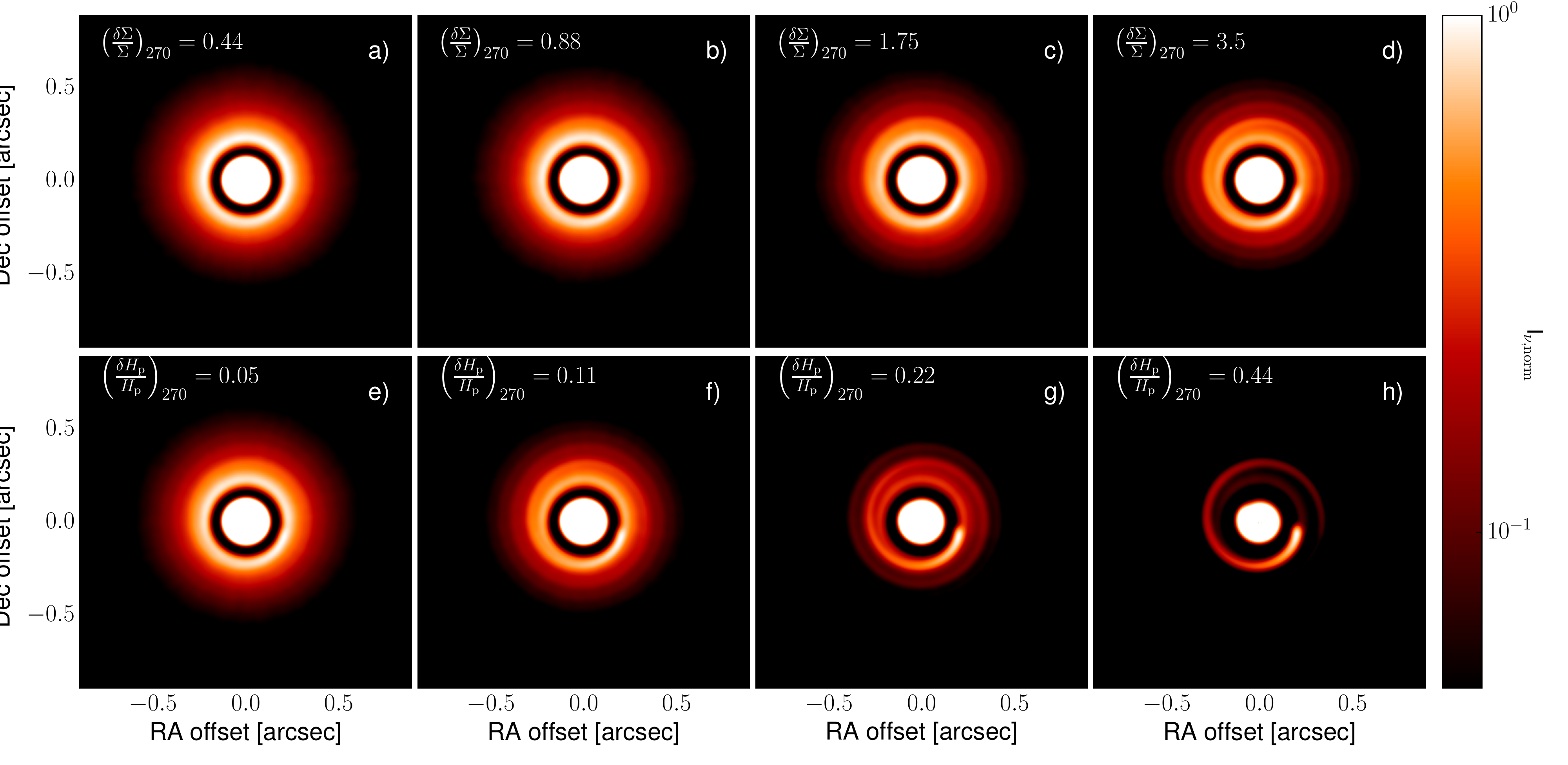}
\caption{Simulated {\it H}-band polarised intensity images with analytic surface density ({\it a}--{\it d}) and pressure scale height  ({\it e}--{\it h}) perturbations.
The images calculated with the radiative transfer code are convolved with a Gaussian PSF with an FWHM of 0.06" and are normalised to the highest 
intensity outside of a 0.18" radius (25.2\,AU at 140\,pc distance) from the central star. The relative change of the perturbed variable at 270$^\circ$ azimuth
angle, i.e. along the negative Dec offset axis, is shown in the top left corner of each panel.
To see the spiral arms in the PSF-convolved images a perturbation amplitude of a factor of 0.22 and 3.5 above the unperturbed disc is required
in the case of pressure scale height and surface density perturbations, respectively.
}
\label{fig:image_analytic_dsigma_beamconv}
\end{figure*}

Eq.\,\ref{eq:pert_eq2}--\ref{eq:sp_wake_muto} describe the 2D disc structure with ten parameters. We fix $\alpha$, $\beta$,$R_{\rm pl}$, $\phi_{\rm pl}$ to the values
we used in the hydrodynamic simulations and we fit $h_{\rm pl}$, $\Sigma_0$,  $p$, $A$, $q$ and $\sigma_{\rm sp}$. We use an eyeball fit to the surface density
distribution and do not use sophisticated optimisation algorithms. The fitted values to the reference hydrodynamic
model are summarised in Table\,\ref{tab:analytic_spiral_fitting}. While value of $h_{\rm pl}$ was also
a known (input) parameter in the hydrodynamic simulation, we need to increase it by 33 per cent in order to match the spiral wake in the hydrodynamic simulation
far away from the planet in the outer disc.  Finally we reduce the surface density by a factor of $10^{-5}$ between 15\,AU and 30\,AU to mimic the 
presence of a gap.

In Fig.\,\ref{fig:FARGO_fit_perturbation}{\it a--c} we compare the fitted analytic surface density distribution to the hydrodynamical
simulations. The analytic model reproduces the structure of the disc, the shape and amplitude of the spirals reasonably well in the outer disc. In the inner
parts the analytic model overestimates the surface density by a factor of 2 (see Fig.\,\ref{fig:FARGO_fit_perturbation}{\it c}). The higher surface density 
in the analytic models may increase the shadowing of the outer disc \citep{juhasz_2007}. Since we are interested only in the contrast of the spiral arms
with respect to the background disc, the increased shadowing by the inner disc would not affect our results as long as the shadowing is axisymmetric. 
We have verified that the spiral perturbation in the inner disc, as described in the analytic model, is too weak and too tightly wound to cause 
any non-axisymmetric shadowing.

We use the above described formalism for the spiral perturbation to calculate a series of models with analytic surface density perturbation.
In reality not only density but also thermal perturbation is expected along the spiral wake. More than a few scale-heights away from the planet shocks 
form along the spiral wake which will introduce also thermal perturbation along the spirals (see e.g.  \citealt{goodman_2001, rafikov_2002a}). 
Since the pressure scale-height of the disc depends on the local temperature the temperature perturbation in turn will cause changes in the vertical
density structure. We study the effect of perturbation in the vertical density structure on the scattered light images by applying the above described analytic 
formalism for the pressure scale-height and run a series of models with various perturbation amplitude.

\begin{figure}
\includegraphics[width=8.0cm]{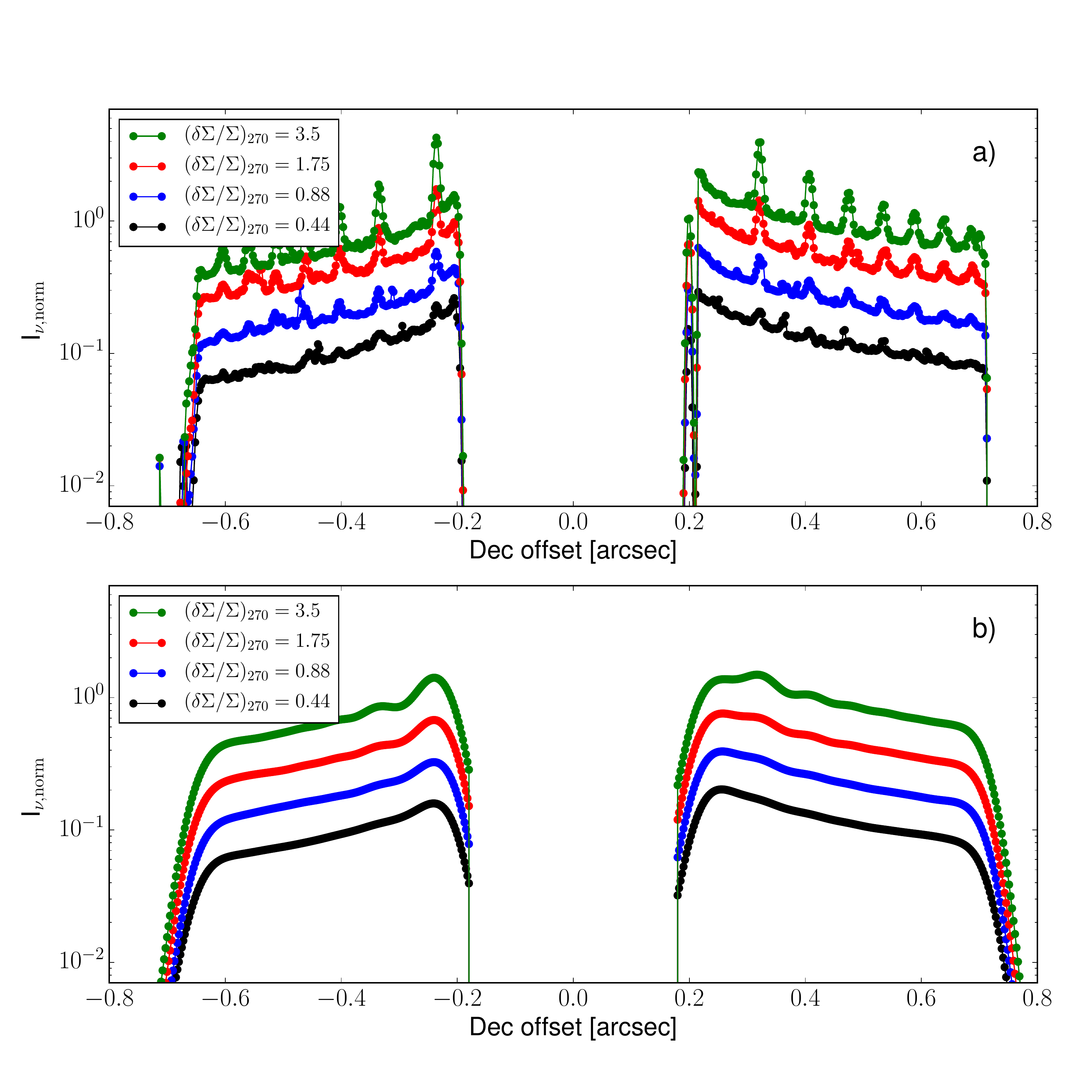}
\caption{Cross-section of the calculated images with different surface density perturbations at original resolution, as calculated by \textsc{radmc-3d} (Panel {\it a}), 
and after convolved with a 0.06" circular Gaussian PSF (Panel {\it b}). }
\label{fig:image_crossection_analytic_dsigma}
\end{figure}

\begin{figure}
\includegraphics[width=8.0cm]{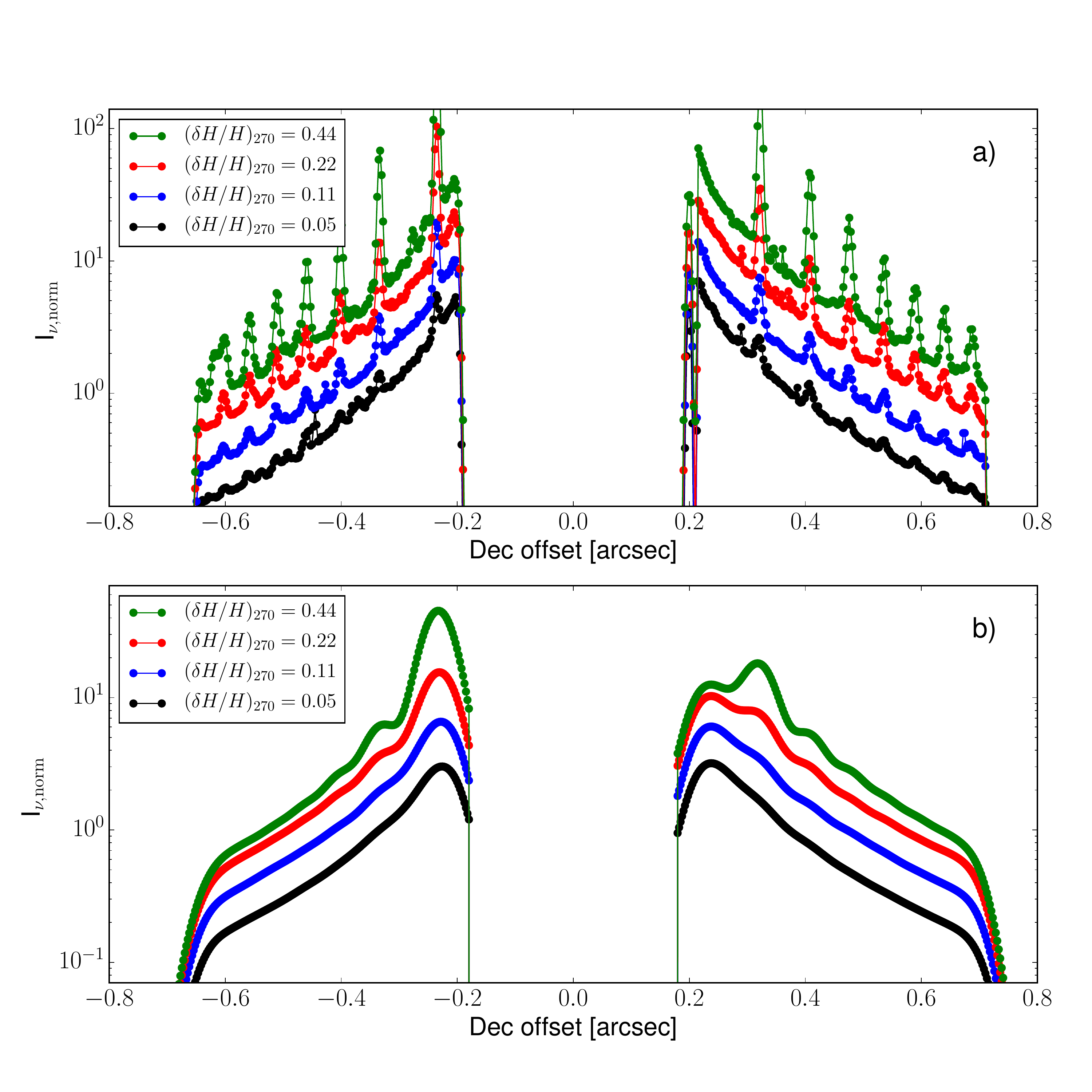}
\caption{Cross-section of the calculated images with different pressure scale height perturbations at original resolution, as calculated by \textsc{radmc-3d} (Panel {\it a}), and after convolved with a 0.06" circular Gaussian PSF (Panel {\it b}). }
\label{fig:image_crossection_analytic_dhp}
\end{figure}

\subsection{Images with analytic perturbations}
\label{subsec:sdperturb_analytic}

\subsubsection{Surface density perturbations}
\label{subsubsec:sdperturb_analytic}
We take the analytic model for the spiral perturbation described in Sec.\,\ref{sec:analytic_models} and whose free parameters are 
fitted to match the shape of the spirals in the hydrodynamic simulations, and change the amplitude of the spiral perturbation by 
changing the value of $A$ in Eq.\,\ref{eq:pert_eq2}. The value of $A$ sets the relative change of the perturbed
variable ($\delta X/X$) at the position of the planet. Massive planets can open a gap where the perturbation may not be visible due to the 
low surface density. Therefore we also calculate the relative change of the perturbed variable  at 270$^\circ$ azimuth
angle from the planet position, at the point where the spiral first crosses the negative y-axis outside of the planetary orbit. We denote
this quantity as $(\delta X/X)_{270}$. In all our models this position is already located in the outer disc and not in the gap, thus the 
spiral can already be observed. For the parameters given in Table\,\ref{tab:analytic_spiral_fitting} the value of $(\delta X/X)_{270}$ is 0.44. 

In Fig.\,\ref{fig:image_analytic_dsigma_beamconv}--\ref{fig:image_crossection_analytic_dsigma} we show the calculated 
{\it H}-band polarised intensity images for the case of surface density perturbations, and the radial cross-section of the images, respectively. It can be seen that the relative change of the surface
density along the spirals needs to be 3.5 at least to make the spirals visible in scattered light images if the FWHM of the PSF is 0.06".
This is a factor 8 times stronger perturbation than what we see in our hydrodynamical simulations for a planet mass of $10^{-3}M_\star$.

\subsubsection{Pressure scale height perturbations}
\label{subsubsec:hpperturb_analytic}

\begin{figure*}
	\centering
	\centerline{
		\includegraphics[width=17cm]{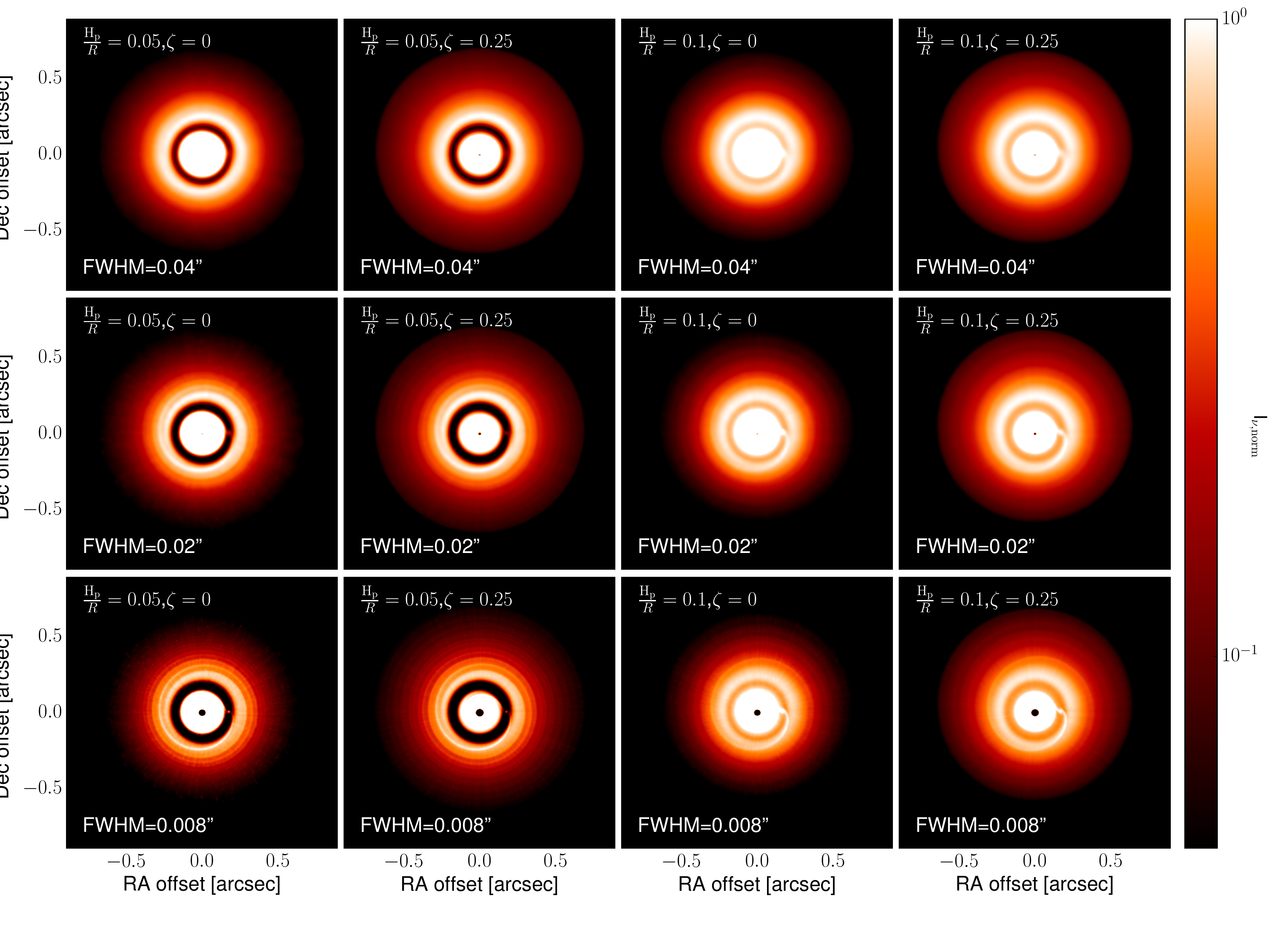}
	}
	\caption{Effect of the spatial resolution on the detectability of spiral arms with pure surface density perturbation. The images are calculated from 
		hydrodynamic simulations for a planet to star mass ratio of 0.001. Panels in different columns show models with different values for the 
		aspect ratio ($H_{p}/R$) and the flaring index ($\zeta$), which is indicated in the top left corner of each panel. The three rows show the
		images convolved with Gaussian PSFs representative in size for SPHERE/VLT in the {\it H}-band (FWHM=0.04", top row), SPHERE/VLT in the {\it R}-band
		(FWHM=0.02", middle row) and E-ELT in the {\it H}-band (FWHM=0.008", bottom row). To detect pure surface density perturbation along the spirals
		at 140\,pc distance one needs an extreme high spatial resolution, where the FWHM of the PSF is 0.02" or smaller. 
				}		
	\label{fig:image_PDI_prediction}
\end{figure*}

We use the same perturbation
function that we used in Sec.\,\ref{subsubsec:sdperturb_analytic} for the surface density, but we apply it to the pressure scale height. 
Similar to the surface density perturbations we  change the strength of the two-dimensional perturbation function to
study the detectability of the spirals.
The resulting images for different values for the relative perturbation amplitude are shown in Fig.\,\ref{fig:image_analytic_dsigma_beamconv}{\it e--h}.
One needs a relative change in the pressure scale height of at least $\sim$0.2 in order to detect the spirals in observations with a PSF whose FHWM is 0.06".

The radial cross-section of the images is presented in Fig.\,\ref{fig:image_crossection_analytic_dhp}. By comparing Figs.\,\ref{fig:image_crossection_analytic_dsigma}-\ref{fig:image_crossection_analytic_dhp}, it can be seen that to achieve a certain contrast between
the spiral features and the background disc we need a factor of about 16--20 smaller relative perturbation in case of  pressure scale height variation, 
than for surface density perturbation.

It is expected that the same relative perturbation of the pressure scale height causes higher density perturbation in the upper layers 
($4H_{p}$ and above) of the disc compared to surface density perturbation. Surface density variations translate to the same
amount of relative perturbations in the volume density of the dust and gas independently of the vertical height above the disc mid-plane. 
In contrast, perturbation in the pressure scale height changes the volume density in the uppermost layers the most. Changes in the
volume density are related to the pressure scale height variations as 

\begin{eqnarray}
\nonumber \frac{\rho(R,z) + \delta\rho(R,z)}{\rho(R,z)} =  \\
\frac{H_{p}(R)}{H_{p}(R) + \delta H_{p}(R)}  \exp{{\left(\frac{z^2}{2H_{p}(R)^2}-\frac{z^2}{2(H_{p}(R) + \delta H_{p}(R))^2} \right)}}
\label{eq:dHp_amp}
\end{eqnarray}

According to Eq.\,\ref{eq:dHp_amp} 20 per cent relative perturbation in the pressure scale height introduces the same volume density
perturbation at the bottom of the disc atmosphere at $4H_{p}$ above the disc mid-plane as relative perturbation of a factor of 
$\sim$8 in the surface density. Since the variation in the surface brightness in scattered light images of protoplanetary discs reflects
volume density variation in the upper layers of the disc, we expect pressure scale height perturbation to be more effective in 
causing observable, i.e. high contrast, perturbations compared to surface density variations. 

\section{Effect of spatial resolution}
\label{subsec:predictions}

The dramatic decrease of the contrast between the spiral features and the background disc is caused by the convolution of the images with the 
telescope PSF.  As mentioned in Sec.\,\ref{subsec:RT_model_assumptions} the images 
calculated by \textsc{radmc-3d} have a resolution of 0.357AU/pixel corresponding to 2.55$\cdot10^{-3}$\,arcsec/pixel, sufficient to resolve the 
spiral arms. These images indeed reflect the contrast between the spirals and the disc we see in the surface density. 
In the PSF convolved images, however, the width of the spiral arms is significantly less than that of the PSF. In this case
the contrast between the spirals and the background disc is lowered by the fraction of the PSF covered by the spiral. 

For this very reason the detectability criterion for the perturbation amplitude depends on the ratio between the width of the spirals
and the size of the PSF. For a source at a given fixed distance from the  observer, the contrast between the disc and the spiral  
increases with increasing resolution as long as the spirals are not spatially resolved. If, however, we fix the resolution, i.e. the size of the
telescope, the contrast between the spiral and the disc increases with decreasing distance to the source. 
In Fig.\,\ref{fig:image_PDI_prediction} we investigated the detectability of  spiral surface density perturbations with next 
generation instruments (SPHERE/VLT) and telescopes (E-ELT).  We take the images, we calculated in Sec\,\ref{subsec:sdperturb} 
based on hydrodynamical simulations of an embedded planet with a mass of 10$^{-3}M_\star$, and convolve them with various 
sized PSFs expected in observations of these instruments. We assume that the source is located at a fixed 140\,pc distance. 

SPHERE on VLT will deliver diffraction limited images at optical and near-infrared wavelengths. In the {\it H}-band the FWHM of the
Airy-disc\footnote{FWHM=$\lambda/D$ radians, where $\lambda$ is the wavelength of observations and $D$ is the diameter of the
primary mirror} in SPHERE observations is $\sim$0.04". This resolution is not yet sufficient to detect pure surface density perturbations
along the spirals in discs at 140\,pc distance (see Fig.\,\ref{fig:image_PDI_prediction}\,top row). One can increase the 
resolution by almost a factor of two if one observes the disc in the {\it R}-band instead of the {\it H}-band.  SPHERE images in the {\it R}-band will have 
high-enough resolution (FWHM=0.02") to detect the spirals (see Fig.\,\ref{fig:image_PDI_prediction}\,middle row).
Since the size of the PSF is still larger than the width of the spirals ($\sim0.013"$) the contrast between the spiral and the disc
is still lower in the scattered light images than in the surface density. Only E-ELT class telescopes can deliver high enough
resolution images (FWHM=0.008") to spatially resolve the spirals (see Fig.\,\ref{fig:image_PDI_prediction}\,bottom row).
E-ELT class telescopes will be able to directly measure the true density contrast between the spirals and the background disc.

\begin{figure*}
\centering
\centerline{\includegraphics[width=17cm]{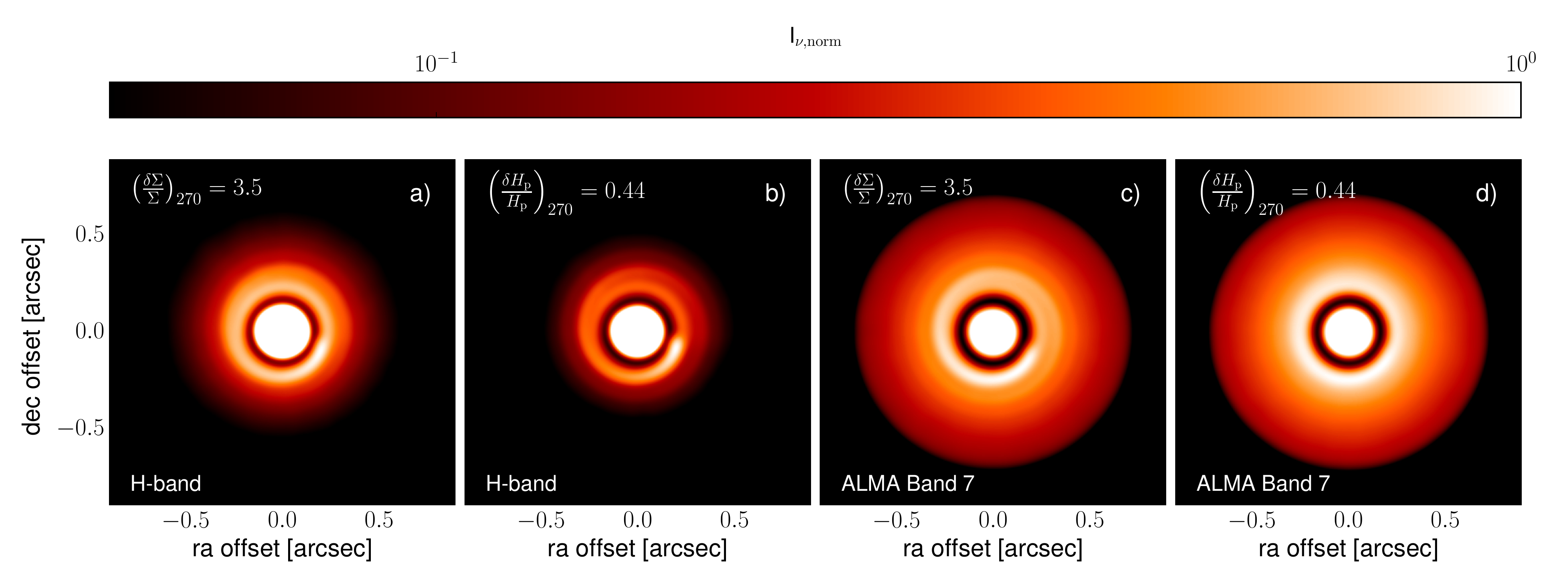}}
\caption{Observability of spirals by surface density (Panel {\it a, c}) and pressure scale height (Panel {\it b, d}) perturbation in near-infrared
polarised intensity (Panel {\it a, b}) and in sub-millimetre thermal emission (Panel {\it c, d})	. The images calculated by the radiative transfer
code are convolved with a circular Gaussian kernel with an FWHM of 60 milli-arcsecond. 
For surface density perturbation the contrast between the spiral and the background disc is the same in the near-infrared scattered light
as in the sub-millimetre. In contrast, pressure scale height perturbation that cause similar amplitude signatures in near-infrared scattered 
have extremely weak, practically unobservable, signature in the sub-millimetre continuum. 
}
\label{fig:image_Hband_ALMAcomp}
\end{figure*}

\section{Multi-wavelength observational diagnostics}
\label{sec:testorigin}

In the previous sections we showed that the amplitude of the surface density perturbations induced by planets along the spirals
is too low to explain the contrast in scattered light images. Thus the observed spirals may represent perturbations
in the upper layers of the disc e.g. by variations of the pressure scale height that changes the density preferentially
high above the mid-plane. Here we discuss observational diagnostics supplementary to scattered light images to constrain the origin of 
the observed spiral arms. 

One possible test is to fit the spiral wake in the scattered light images and the SED {\it simultaneously}. The pitch angle of the spirals
is determined by the sound-speed (i.e. temperature) distribution in the disc, which in turn sets the pressure scale height.  
Thus by fitting the spiral wake we can constrain the pressure scale height, i.e. the vertical structure of the disc, as a function of radius.
The scale height and its variation as a function of radius (i.e. the flaring index) determines also the luminosity of the disc and the slope of the
SED from mid- to far-infrared wavelengths, respectively. If the observed spiral arms are caused by planets there must be a solution for
the pressure scale height and sound speed distribution that fits both the SED and the spiral wake in the scattered light images. 
The lack of such solution would imply that the spirals are not caused by planetary mass companions in locally isothermal gravitationally
stable discs. 

There are also methods to test observationally whether the spiral arms are perturbation in the surface density or only confined to the upper
layers of the disc, by comparing sub-millimetre continuum images to those in scattered light. 
At sub-millimetre wavelengths we see the thermal emission of the dust. If the disc is optically thin, which it is 
in most cases outside of about 5--10\,AU\footnote{The disc might become optically thick in sub-millimetre continuum in dust traps
in the cores of anticyclonic vortices.} the sub-millimetre emission probes the total column density of the dust. If the spirals observed in scattered
light are caused by surface density perturbation one should observe the same contrast between the spirals and the background disc at both
wavelengths. 

In contrast to surface density perturbation, scale-height perturbation does not change the column density of the dust and gas, it merely
re-distributes material along the vertical direction. However, scale-height variations are linked to temperature variation and we may
observe variation of the temperature along the spirals. Since $H_{\rm p}\propto\sqrt(T)$ the temperature variation is
$(1+\delta T/T) = (1+\delta H_{\rm p}/H_{\rm p})^2$. This temperature perturbation will be identical to the variation of the flux at sub-millimetre
wavelengths due to the linear dependence of the flux on the temperature in the Rayleigh-Jeans approximation. As we showed in Sec.
\ref{subsubsec:hpperturb_analytic} a given contrast in scattered light images between the spirals and the disc requires a much
lower scale height / temperature variation than surface density perturbation. Thus, in case of pressure
scale height perturbation we expect to see a significantly lower contrast between the spirals and the background disc in the 
sub-millimetre continuum than in near-infrared polarised intensity. 
Note, that the images in sub-millimetre and near-infrared should have the same angular resolution for a straightforward comparison.

To demonstrate this test we take the models with analytical surface density and pressure scale height perturbations from
Sec.\ref{subsubsec:sdperturb_analytic} and Sec.\ref{subsubsec:hpperturb_analytic}, respectively, and calculated images at 880\,$\mu$m, 
representative for ALMA Band 7 observations. We choose the amplitude of the perturbation such that both perturbation type (surface density 
and scale height) should cause a similar contrast signature in the near infrared. For the model with pressure scale height variation we 
changed (i.e. increased) the dust temperature calculated by RADMC-3D by $(1+\delta H_{\rm p}/H_{\rm p})^2$ along the spirals before 
raytracing to simulate the thermal perturbation required for the imposed scale height variations. We neglected the thermal perturbation
in the calculation of the near-infrared images, as thermal emission is negligible in the near-infrared at several tens of AUs distance from
the central star. Then we convolve both the near-infrared polarised intensity images and 
the sub-millimetre images with a 2D Gaussian kernel with an FWHM of 0.06". We note that 0.06" angular resolution can already  be achieved
with ALMA in Cycle 3. The resulting images are presented in Figure\,\ref{fig:image_Hband_ALMAcomp}.  As can be seen the contrast in the 
scattered light and sub-millimetre thermal emission is the same for the surface density perturbation. In case of pressure scale-height variation 
the contrast between the spirals and the disc in the sub-millimetre is so low that the spirals are barely visible. 

\section{Summary and conclusions}
\label{sec:conclusions}

We investigated whether or not the spiral arms seen in protoplanetary discs in polarised scattered light 
\citep{muto_2012, grady_2013, avenhaus_2014a, avenhaus_2014b} can be interpreted as density waves launched 
by one or more planets under the assumption that the disc is locally isothermal. 
There are several important properties of the observed spirals that we can compare to the
properties of planet-driven spiral density waves, i) the number of spiral arms observed and their relative strength, 
ii) the pitch angle of the spiral and iii) the contrast between the spiral and the background disc.

We use 2D hydrodynamic simulations to investigate the morphology of the spirals driven
by single planets. We find that two or even more armed spirals can be excited by a single planet. The number of 
spirals increases for higher planet masses and for lower aspect ratios. These spirals are not symmetric, though, as they have different 
amplitude and width. More than one spiral arm is only seen in simulations where the planet is able to open a well defined
deep gap. Thus to explain the presence of the {\it symmetric} two armed spiral seen in HD135344B \citep{muto_2012} two companions 
are required. Whether or not such configuration (combination of planet masses and orbital parameters) is dynamically stable,  
how long it can be sustained, i.e. how likely it is to observe such system, and whether or not the amplitude of the spirals could be reproduced
with the two companions is yet to be investigated. 

All spirals seen in scattered light observations seems to be open, i.e. the pitch angle seems to be large, which bears important consequences 
for the vertical structure of the disc. On one hand, the more open the spiral the higher the aspect ratio of the disc needs to be and also the smaller
the amplitude of the spiral will become. On the other hand the aspect ratio of the disc, i.e. the pressure scale height and its radial variation, sets the 
amount of reprocessed stellar radiation, which in turn will determine the luminosity of the disc, the shape of the SED from near- to far-infrared 
wavelengths as well as the absolute surface brightness of the disc. Thus {\it simultaneous} fitting of the spiral wake and the SED with a single 
$H_{\rm p}(R)$ curve can be a powerful tool to test whether or not the spirals are driven by an embedded planet.

Our simulations show that planets by themselves in gravitationally stable locally isothermal discs cannot create strong enough surface density
perturbation along the spirals that could be observed in near-infrared scattered light with current 8\,m class telescopes. 
Planets in such discs can create a relative surface density perturbation on the
order of several tens of percent at most. This contrast is however lowered to below the detection limit by the convolution with 
the telescope PSF, which is significantly larger than the width of the spirals. We use analytic models to estimate that
a surface density perturbation of a factor of 3.5 or higher above the background disc is required to create spiral density waves 
which are observable at a spatial resolution of 0.06". Such high amplitude perturbation is unlikely to be caused by planetary mass 
companions.

We also test the effect of pressure scale height perturbation along the spirals to the scattered light images. The variation
in the volume density, caused by pressure scale height perturbations, increases monotonically with height above two scale heights. 
One needs a factor of $\sim$16--20 lower relative change in the scale height to create
the same contrast signature in the scattered light images as any given surface density perturbation. In order for the spirals
to be detectable in images convolved with a PSF with an FWHM of 0.06" a relative change in the pressure scale height of $\sim$0.2
is required above the background disc. We therefore suggest that the spiral arms observed so far in protoplanetary discs 
are caused by variations in the vertical structure of the disc (e.g. pressure scale height perturbation) instead of surface density perturbation.

We also investigate the effect of spatial resolution on the detectability of surface density perturbations along the spirals. 
We predict that spiral surface density perturbations cannot be detected with current 8\,m class telescopes in the {\it H}-band, 
even with state-of-the art high-contrast imagers, such as SPHERE on VLT. While the detection of planet-induced spiral surface
density perturbation is possible in the {\it R}-band with 8\,m-class telescopes, an E-ELT class telescope is required to fully resolve
the spirals at near-infrared wavelengths and measure the true density contrast between the spiral and the background disc.

Finally we suggest two types of tests to study the origin of the spirals. To study whether the spirals seen in scattered light  
represent surface density perturbation or scale height variation we can compare sub-millimetre and scattered light images.
In case of surface density perturbation the variation of the density is independent of the height above the disc mid-plane.
Therefore the observed contrast between the spiral and the disc should be the same in sub-millimetre as in scattered light
if the spatial resolution is the same. For scale height variation we should observe a significantly lower contrast in sub-millimetre
as in scattered light due to the lower amplitude of the thermal perturbation required to produce the observed scattered light
signatures. For planet-induced spirals the pitch angle of the spirals is determined by the sound-speed distribution in the disk.
The sound-speed depends on the temperature, which in turn will determine the scale height and the flaring index of the disc.
The flaring index and pressure scale height can be constrained by fitting the SED of the disc, while the pitch
angle can be derived by fitting the spiral wake in the scattered light images. Thus a simultaneous fitting of the SED and the
spiral wake can help to understand whether planets can be responsible for the observed spirals and whether spiral arms
in protoplanetary discs are indeed a signposts of on-going planet formation.

\section*{Acknowledgments}
We are grateful to Paola Pinilla, Cathie Clarke and Giovanni Rosotti for inspiring discussions and comments on the manuscript.
This work has been supported by the DISCSIM project, grant
agreement 341137 funded by the European Research Council under
ERC-2013-ADG. We thank the anonymous referee for the insightful comments that helped to improve our paper. 

\bibliographystyle{mn2e}
\bibliography{ms}

\appendix

\bsp

\label{lastpage}

\end{document}